\documentclass[aps,prl,twocolumn,10pt,groupedaddress,nofootinbib,superscriptaddress,notitlepage,showpacs,floatfix]{revtex4-1}
\usepackage{amsthm,amsmath,amssymb,amsfonts,mathrsfs,bm,braket}
\usepackage{graphicx,graphics,subfigure}
\usepackage{dcolumn,array}
\usepackage{enumitem}
\usepackage{url}
\usepackage[pdfstartview=FitH,pdfencoding=auto]{hyperref}
\usepackage{bookmark}
\usepackage{pifont}
\newcommand{\defeq}{:=}           			% be defined as
\begin{document}
\title{Complementary relation of quantum coherence and quantum correlations in multiple measurements}

\author{Zeyang Fan}
\affiliation{School of Astronautics, Harbin Institute of Technology, Harbin 150001, China}

\author{Yi Peng}
\affiliation{Beijing National Laboratory for Condensed Matter Physics, Institute of Physics, Chinese Academy of Sciences, Beijing 100190, China}
\affiliation{School of Physical Sciences, University of Chinese Academy of Sciences, Beijing 100190, China}

\author{Yu-Ran Zhang}
\affiliation{Beijing Computational Science Research Center, Beijing 100094, China}
\affiliation{Theoretical Quantum Physics Laboratory, RIKEN Cluster for Pioneering Research, Wako-shi, Saitama 351-0198, Japan}

\author{Shang Liu}
\affiliation{School of Physics, Peking University, Beijing 100871, China}

\author{Liang-Zhu Mu}
\email{muliangzhu@pku.edu.cn}
\affiliation{School of Physics, Peking University, Beijing 100871, China}
\author{H. Fan}
\email{hfan@iphy.ac.cn}
\affiliation{Beijing National Laboratory for Condensed Matter Physics, Institute of Physics, Chinese Academy of Sciences, Beijing 100190, China}
\affiliation{CAS Center for Excellence in Topological Quantum Computation,
University of Chinese Academy of Sciences, Beijing 100190, China}

%\date{\today}
%\pacs{ }
\begin{abstract}
Quantum coherence and quantum correlations lie in the center of quantum information science, since they
both are considered as fundamental reasons for significant features of quantum mechanics different from
classical mechanics. We present a group of complementary relations for quantum coherence and quantum
correlations; specifically, we focus on thermal discord and conditional information in scenarios of multiple measurements. We show that the summation of quantum coherence quantified in different bases has a
lower bound, resulting from entropic uncertainty relations with multiple measurements. Similar results are
also obtained for thermal discord and for post-measurement
conditional information with multiple measurements in a multipartite system. These results indicate the
general applications of the uncertainty principle to various concepts of quantum information.
\end{abstract}
\maketitle

Quantum coherence, namely a principle of superposition of quantum states, is one of the cornerstones of
the quantum theory. It is believed to provide advantages in tasks of quantum information processing
and quantum computation over classical methods. Recently, the rigorous framework of the quantification of
quantum coherence has been introduced with several criteria proposed for any coherence measure to
satisfy \cite{baumgratz2014quantifying,Added0A,du2015conditions,shao2015fidelity,yaun2015intrinsic,Lostaglio15,zhang2016quantifying,xu2016quantifying,peng2016maximal,
winter2016operational,Bu2016catalytic,rana2016trace,Added1G,Added2J,Added0N,Chitambar16,Brandner17}. Based on this framework, different quantitative studies of quantum coherence have also
attracted many attentions \cite{bromley2015frozen,chen2015coherence,bera2015duality,hillery2016coherence,cheng2015complementary,marvian2016quantum,bagan2016relations,yu2016measure,misra2016energy},
see reviews \cite{Review1,Review2,Review3} for progresses and references.
Furthermore, quantum coherence is closely related to various quantum correlations, such as entanglement
and quantum discord \cite{streltsov2015measuring,ma2016converting,killoran2016converting,chitambar2016relating}.
In this paper, we focus on two well known quantum correlations: thermal discord \cite{zurek2003quantum,horodecki2005local,modi2012classical} and conditional information \cite{nielsen2010quantum}. Also known as one-way deficit, thermal discord is often studied in the quantum thermodynamics, e.g., the research of Maxwell's Demon \cite{horodecki2005local,modi2012classical}.
Conditional information plays an important role in quantum entanglement \cite{cerf1997negative,cerf1999quantum,plenio2007introduction} and state merging \cite{horodecki2005partial,horodecki2007quantum}, because a negative value of conditional information signals
the existence of entanglement.
Apart from these quantum quantities, the Heisenberg uncertainty principle \cite{heisenberg1927anschaulichen,robertson1929uncertainty,Ozawa,deutsch1983uncertainty,kraus1987complementary,maassen1988generalized,berta2010uncertainty,liu2015entropic,coles2012uncertainty}
has been widely studied in the quantum information processing. These studies, including entropic
complementary relations \cite{deutsch1983uncertainty,kraus1987complementary,maassen1988generalized,berta2010uncertainty,liu2015entropic,coles2012uncertainty}, not only provide a deeper understanding of  quantum mechanics, but also give a useful tools for researches on the quantum information processing.

In sharp contrast to quantum entanglement that is invariant for any local unitary transformation, the
quantification of quantum coherence is in general basis dependent and depends on both the quantum state
itself and the basis we choose. It is then an interesting question whether we can find a class of basis-independent
coherence measures or diminish the effects of the basis chosen. Based on this consideration, in this work,
we investigate the total quantum coherence for multiple bases and present the results in forms of
complementary relations. We start from uncertainty relations \cite{berta2010uncertainty,liu2015entropic} with measurements on different subsystems and multiple measurements. We obtain a group of complementary
relations for quantum coherence, thermal discord and conditional information, which give a deeper understanding
of these concepts. Explicitly, complementary constraints are given on quantum coherence and basis-dependent
thermal discord with respect to multiple measurement bases. As an example, we investigate quantum coherence
in a one-particle system and thermal discord in a bipartite system by implementing all the measurements on a
single subsystem. In addition, for multiple measurements in a multipartite system,
we also present a complementary relation for post-measurement conditional
information.

\section{Results}	
\subsection{Coherence quantification and the definitions of thermal discord and conditional information}
A natural measure of quantum coherence is defined as a pseudo-distance formulated by the relative entropy
between the studied quantum state with the nearest incoherent state. It can be proven that this nearest
incoherent state is the corresponding diagonal matrix of the studied density matrix with all off-diagonal
elements zero \cite{baumgratz2014quantifying},
	\begin{equation}
		C_\textrm{RE}^{(\mathcal{M})}(\rho ) = S(\tilde {\rho}^{(\mathcal{M})}) - S\left(\rho\right),
		\label{coherence_re}
	\end{equation}
where $S\left(\rho\right){\defeq} -{\rm Tr}\rho \log _2\rho $ is the von Neumann entropy,
$\mathcal{M}{\defeq}\{\Pi_i{\defeq}\ket{i}\bra{i}\}$ is the projective measurement, and
	$\tilde {\rho}^{(\mathcal{M})}{\defeq}\sum_i\Pi_i\rho \Pi_i$ is the post-measurement state.
It is apparent that the post-measurement state $\tilde {\rho}^{(\mathcal{M})}$ is the diagonal matrix
of the density matrix $\rho $, $\tilde {\rho}^{(\mathcal{M})}=\rho _{\rm diag.}$. In this sense, the relative
entropy of coherence in Eq.(\ref{coherence_re}) can be understood as the increase of the entropy via a projective
measurement $\mathcal{M}$, meaning the difference of the entropy between the post-measurement state and the original state.
Therefore, quantum coherence describes the quantum resource destroyed in the projective
measurement, which is naturally interpreted from the projective-measurement point of view.
Because quantum coherence measures such as $C_\textrm{RE}^{(\mathcal{M})}(\cdot)$ are basis dependent,
finding the relation between quantum coherence of the same state with respect to different measurement bases
$\mathcal{M}$ would be significant. By varying the measurement basis, quantum coherence changes and has
a maximum \cite{huml}. Then, considering multiple measurements, it is not clear what is the relationship
among those quantifications of quantum coherence based on different bases. We will derive those relations
using entropic uncertainty inequalities for multiple measurements with and without the quantum memory.

Let us consider a bipartite state $\rho _{\rm AB}$ with two subsystems A and B. The projective measurement
$\mathcal{M}^{(k)}_{\rm A}{\defeq}\{\Pi_{{\rm A}i}^{(k)}{\defeq}\ket{i^{(k)}}_A\bra{i^{(k)}}\}$ is performed on the
subsystem A, and we have the unnormalized post-measurement state as
\begin{eqnarray}
p_i^{(k)}\tilde{\rho}_{\textrm{AB}i}^{(k)}&=&(\Pi_{{\rm A}i}^{(k)}\otimes I)\rho _{\rm AB}(\Pi_{{\rm A}i}^{(k)}\otimes I)
\nonumber \\
&=&p_i^{(k)}\ket{i^{(k)}}_{\rm A}\bra{i^{(k)}}\otimes \tilde{\rho}_{\textrm{B}i}^{(k)},
\label{measureA}
\end{eqnarray}
where $k$ labels the reference basis, and the probability of obtaining a result $i$ is
$p_i^{(k)}={\rm Tr_{AB}}(\Pi_{{\rm A}i}^{(k)}\otimes I)\rho _{\rm AB}(\Pi_{{\rm A}i}^{(k)}\otimes I)$.
Then, thermal discord is defined as~\cite{zurek2003quantum,modi2012classical}
	\begin{equation}
		D^{(k)}_\textrm{th}(\textrm{B}|\textrm{A})
		{\defeq}\sum_{i=1}^dp_i^{(k)} S(\tilde{\rho}_{\textrm{B}i}^{(k)}) + S(\tilde{\rho}_\textrm{A}^{(k)}) - S(\rho_\textrm{AB}).
		\label{thermalDiscord_def}
	\end{equation}
with $\tilde{\rho}_{\textrm{B}i}^{(k)}={\rm Tr_A}(\Pi_{{\rm A}i}^{(k)}\otimes I)\rho _{\rm AB}(\Pi_{{\rm A}i}^{(k)}\otimes I)/p_i^{(k)}$
the post-measure state of subsystem B with probability $p_i^{(k)}$ and
\begin{eqnarray}
\tilde{\rho}_{\textrm{A}}^{(k)}&=&{\rm Tr_{\rm B}}\left[\sum _i(\Pi_{{\rm A}i}^{(k)}\otimes I)\rho _{AB}(\Pi_{{\rm A}i}^{(k)}\otimes I)\right]
\nonumber \\
&=&\sum _ip_i^{(k)}\ket{i^{(k)}}_{\rm A}\bra{i^{(k)}},
\end{eqnarray}
the state of the subsystem A after the measurement without knowing any outcome. Thermal discord concerns
about the entropic cost of performing a local  projective measurement on the subsystem of a bipartite
state and is relevant with the thermodynamics of correlated systems \cite{modi2012classical}. The definition of
thermal discord is also measurement-dependent. We can generally make a measurement-independent definition
by requiring a minimization over all projective measurements, but this would in principle complicate the calculation
and make it difficult to obtain a closed expression. In the experiment, we would always choose specific
measurements when the optimal measurement is not available. In the meantime to gain more information, one
would implement more than one measurements with different bases. In such a situation, it would be very helpful
if we know any relation among  thermal discord with multiple measurements.
	
Moreover, conditional information on joint system $\textrm{A}\textrm{B}$ is defined as
	$S\left(\textrm{A}|\textrm{B}\right){\defeq}S\left(\rho_{\textrm{A}\textrm{B}}\right)-S\left(\rho_{\textrm{B}}\right)$~\cite{nielsen2010quantum}.  We denote  the conditional
	entropy after a measurement $\mathcal{M}_\textrm{A}$ on the subsystem A as
	$S\left(\mathcal{M}_\textrm{A}|\textrm{B}\right)=S\left(\tilde{\rho}_\textrm{AB}\right)-S\left(\tilde{\rho}_\textrm{B}\right)$.
In this case, instead of multiple measurements on one system, we may have a complementary
	relation on post-measurement conditional information in a multipartite system.

\subsection{Complementary relation of quantum coherence}
The corresponding post-measurement state for a measurement $\mathcal{M}^{(k)}$  and a quantum state $\rho $ can be written as,
 \begin{eqnarray}
 \tilde {\rho}^{(k)}=\sum _i\Pi_i^{(k)}\rho \Pi_i^{(k)}= \sum _ip_i^{(k)}\tilde {\rho }_i^{(k)},
 \end{eqnarray}
which can be written in a more explicit form as,  $\tilde {\rho}^{(k)}=\sum _ip_i^{(k)}\ket{i^{(k)}}\bra{i^{(k)}}$.
 The entropy of this post-measurement state takes the form,
 \begin{eqnarray}
 S(\tilde {\rho}^{(k)})=-{\rm Tr}\tilde {\rho}^{(k)}\log _2\tilde {\rho}^{(k)}=
 -\sum _ip_i^{(k)}\log_2p_i^{(k)},
 \label{2addmeasure}
 \end{eqnarray}
which describes how exactly we can measure the state $\rho $ by the measurement operator $\mathcal{M}^{(k)}$.
 For example, for a projective measurement in computational basis $\{|0\rangle ,|1\rangle \}$ on a qubit state,
 if we obtain the measurement results $|0\rangle $ and $|1\rangle $ with an equal probability $1/2$,
 the entropy in Eq.~(\ref{2addmeasure}) is 1. In this case, we do not know whether the state should be
 $|0\rangle $ or $|1\rangle $, since both the probabilities to obtain those two states are $1/2$.
 The state corresponds to a completely mixed
 state or a maximally coherent state $|+\rangle $ or $|-\rangle $, where $|\pm \rangle =(|0\rangle +|1\rangle )/\sqrt {2}$.
 If we obtain $|0\rangle $ or $|1\rangle $ with probability 1, the result of relation (\ref{2addmeasure}) is 0. We know exactly whether the state is
 $|0\rangle $ or $|1\rangle $. The principle of uncertainty in quantum mechanics states that we cannot achieve
 arbitrary measurement precision simultaneously for non-commuting observables, meaning that the summation of  entropies for
 the post-measurement states for non-commuting observables should have a lower bound larger than 0 for an arbitrary quantum state.
 Generally, the entropic uncertainty relation with $N$ measurements
 $\{ \mathcal{M}^{(k)}\}_{k=1}^N$ can be written as follows \cite{liu2015entropic}
	\begin{eqnarray}
		\sum _{k=1}^N S(\tilde {\rho}^{(k)}) \ge -\log _2b+(N-1)S(\rho ),
		\label{uncertainty}
	\end{eqnarray}
	where $b\in(0,1]$ is  determined by measurement operators $\{ \mathcal{M}^{(k)}\}_{k=1}^N$,
	\begin{eqnarray}
		b&=&\max_{i_N}\left\lbrack\sum_{i_2,\ldots,i_{N-1}}\max_{i_1}[{{\rm Tr} (\Pi^{(1)}_{i_1},\Pi^{(2)}_{i_2})}]\right.	\nonumber\\
		 &&\qquad\qquad\qquad\left. \times\prod_{k=2}^{N-1}{\rm Tr} (\Pi^{(k)}_{i_k},\Pi^{(k+1)}_{i_{k+1}})\right\rbrack .
		\label{Liubound}
	\end{eqnarray}
Here, $\tilde{\rho}^{(k)}$ is the post-measurement state for $\mathcal{M}^{(k)}$, as we have mentioned, where
${\rm Tr} (\Pi^{(1)}_{i_1},\Pi^{(2)}_{i_2})=|\langle i_1^{(1)}|i_2^{(2)}\rangle |^2$ corresponds to the square of the
overlap between two different bases. We remark that quantity $b$ depends only on the measurement set
$\{ \mathcal{M}^{(k)}\}_{k=1}^N$, and no order should be a priori assumed for
$\{ \mathcal{M}^{(k)}\}_{k=1}^N$ in Eq.~(\ref{Liubound}), so $b$ is state independent. We adjust
(\ref{uncertainty}) to another form
	\begin{equation}
		\sum_{k=1}^{N}\left\lbrack{S(\tilde{\rho}^{(k)}) - S(\rho)}\right\rbrack
		{\ge} -\log _2{b} -S\left(\rho\right)
	\end{equation}
	for  ease of obtaining the complementary relation of quantum coherence. Since the relative entropy coherence measure $C_\textrm{RE}^{(k)}(\rho)$ corresponding to the
	reference basis  $\lbrace{\Pi_i^{(k)}}\rbrace$ can be defined as the increase of the entropy $S(\rho^{(k)}) - S(\rho)$ of the system due to
	measurement $\mathcal{M}^{(k)}$, we  can equivalently obtain
	\begin{equation}
		 \sum_{k=1}^{N}C_\textrm{RE}^{(k)}\left(\rho\right)
		 {\ge} -\log_2{b} - S\left(\rho\right),
		 \label{CR_RE-Coherence}
	\end{equation}
which provides a lower bound for the relative entropy of quantum coherence in addition to the upper bound given in Ref~\cite{cheng2015complementary}. In particular, if the state is pure, $S(\rho)=0$, the bound shown in the
right-hand-side of the inequality (\ref{CR_RE-Coherence}) depends only on measurement bases, and can be
larger than that of a mixed state. In this case, the bound can be a finite positive value if we choose different
measurement bases, implying that the quantum coherence of a pure state can always be non-zero if we can
choose an appropriate basis. We remark that the inequality (\ref{CR_RE-Coherence}) is simply a different form
of the entropic uncertainty relation with multiple measurements \cite{liu2015entropic}. However, the implication
of this inequality is different as the inequality (\ref{CR_RE-Coherence}) is for total coherence with different bases. In particular, when there is no fixed basis in measuring
coherence, a constraint in combining quantum coherence measured in different bases will be insightful. The complementary relation
describes the properties of coherence from this point of view.

We know that the relative entropy of coherence $C_\textrm{RE}^{(k)}\left(\rho\right)$ is
	non-negative, however, the right-hand-side of the inequality can be positive, zero or even negative. In particular, when $\rho $ is a mixed state, $S(\rho )$ can be relatively large,
	so the lower bound will probably be negative. This fact is reasonable because that coherence depends on the chosen basis, i.e., projective measurement $\mathcal{M}$. Let us consider
	an extreme case when $\rho $ is a completely mixed state, $S(\rho )=\log _2d$, where $d$ is the dimension of Hilbert space. The coherence is always zero regardless of the basis,
	so the total coherence in the left-hand-side of Eq.~(\ref{CR_RE-Coherence}) is also zero, while right-hand-side is no larger than 0, $-\log _2b-S(\rho )\leq 0$. Then, the inequality is always true for
	arbitrary sets of $\{\mathcal{M}^{(k)}\}_{k=1}^N$. The equality can be satisfied for some specifically bases such as the mutually unbiased bases.
This fact also indicates that we can expect a higher bound if the state has a larger purity resulting in a smaller
	entropy. We notice that this
	result (\ref{CR_RE-Coherence}) was also reported in Ref.~\cite{singh2016uncertainty} very recently.

For example, let us study a density matrix as follows,
\begin{eqnarray}
\rho =p|\psi \rangle \langle \psi |+\frac {1-p}{2}{\textbf{I}},
\label{example}
\end{eqnarray}
which is constituted by a maximally coherent state $|\psi \rangle $ and a completely mixed state
${\textbf{I}}/2$, where $p\in [0,1]$. This maximally coherent state takes the form,
\begin{eqnarray}
|\psi \rangle =\frac {1}{\sqrt {2}}(|0\rangle +e^{i\phi }|1\rangle ),
\end{eqnarray}
where $\phi \in [0,2\pi )$ is the phase parameter, and ${\textbf{I}}$ is the identity operator in two-dimensional Hilbert space.
We consider two measurement sets $\{|0\rangle ,|1\rangle \}$ and $\{|+\rangle ,|-\rangle \}$, respectively.
In this case, $b=\frac {1}{2}$, based on Eq.(\ref{Liubound}). The entropy of state $\rho $ is,
\begin{eqnarray}
S(\rho )=-\frac {1+p}{2}\log _2\frac {1+p}{2}-\frac {1-p}{2}\log _2\frac {1-p}{2}.
\label{entropy-p}
\end{eqnarray}
The diagonal form of the density matrix can be written respectively in two bases, $\{|0\rangle ,|1\rangle \}$, $\{|+\rangle ,|-\rangle \}$,
as follows,
\begin{eqnarray}
\rho _{diag.}&=&\left\{ \frac {1}{2}, \frac {1}{2}\right\} _{\{ |0\rangle ,|1\rangle \}} ,\\
\rho _{diag.}'&=&\left\{ \frac {1+p\cos \phi }{2}, \frac {1-p\cos \phi }{2}\right\} _{\{ |+\rangle ,|-\rangle \}}.
\end{eqnarray}
Thus, for the computational basis, the entropy of the diagonal density matrix is
$S(\rho _{diag.})=1$. For the basis $\{|+\rangle ,|-\rangle \}$, the
entropy of the diagonal density matrix is,
\begin{align}
S(\rho _{diag.}')=-\frac {1+p'}{2}\log _2\frac {1+p'}{2}
-\frac {1-p'}{2}\log _2\frac {1-p'}{2},
\label{entropy-p1}
\end{align}
where we denote $p'=p\cos \phi $ for convenience.

By using Eq.(\ref{uncertainty}) and Eq.(\ref{CR_RE-Coherence}),  the complementary relation for coherence
implies the following inequality,
\begin{eqnarray}
\left[ S(\rho _{diag.})-S(\rho )\right] +\left[ S(\rho _{diag.}')-S(\rho )\right]\ge -\log _2b-S(\rho ).
\nonumber \\
\end{eqnarray}
The bound in right-hand-side takes the value, $1-S(\rho )$, meaning that the total coherence of state $\rho $
based on two measurement bases should be larger than a positive value unless it becomes a completely mixed state
which leads to $S(\rho )=1$. On the other hand, for a pure state $\rho $, $S(\rho )=0$, the bound equals to 1.
So the total coherence should be always larger or equal to 1.
Thus, we know that coherence of state $\rho $ as a resource depends not only on basis,
but also on the von Neumann entropy of the state.

Next, we show explicitly that this inequality is true.
Substituting the results, $S(\rho _{diag.})=1$ and $b=\frac {1}{2}$, the inequality becomes,
\begin{eqnarray}
S(\rho _{diag.}')-S(\rho )\ge 0.
\end{eqnarray}
By considering the results in Eq.~(\ref{entropy-p}) and Eq.~(\ref{entropy-p1}), we know this inequality
is correct based on the fact $p'\le p$. When $\phi =0$ meaning $p=p'$, the inequality becomes  an equality.
This result can also be understood as the fact that the relative entropy of coherence for state $\rho $ is nonnegative.

Although the quantum coherence measure is basis dependent, for multiple measurement bases, the summation of
the coherence can be bounded from below by a quantity depending on $b$ determined by the overlap between different bases and
the entropy of the state. The parameter $b$ itself is independent of the studied state.
So there exists the complementary relation for coherence with multiple measurements.

\subsection{Complementary relations of thermal discord}

For the quantum state $\rho _{\rm AB}$, in case the conditional entropy, also in name of conditional information, is
negative, $S\left(\textrm{A}|\textrm{B}\right){\defeq}S\left(\rho_{\textrm{A}\textrm{B}}\right)-S\left(\rho_{\textrm{B}}\right)<0$,
we know that there is entanglement between A and B. In this case, suppose that subsystem
B plays the role of quantum memory, the uncertainty extent of subsystem A for measurements will decrease \cite{berta2010uncertainty}.
In this physical setting, we perform multiple measurements on A, the multi-measurement uncertainty
with the assistance of memory B can be derived as follows \cite{liu2015entropic}
	\begin{equation}
		\sum_{k=1}^NS(\mathcal{M}^{(k)}_\textrm{A}|\textrm{B})
		{\ge} - \log _2{b} + (N-1)S(\textrm{A}|\textrm{B})
		\label{Multi-Meas_UP_Memory}.
	\end{equation}
where $S(\mathcal{M}^{(k)}_\textrm{A}|\textrm{B})=
S(\tilde {\rho }_{\rm AB}^{(k)})-S(\tilde {\rho }_{\rm B}^{(k)})$
is the conditional information after the measurement $\mathcal{M}^{(k)}_\textrm{A}$ on A,
  recalling the notations and results in Eq.~(\ref{measureA}), and we have $\tilde {\rho }_{AB}^{(k)}=\sum _ip_i^{(k)}\tilde{\rho}_{\textrm{AB}i}^{(k)}$.
  We can rewrite the first
	term of thermal discord in the definition (\ref{thermalDiscord_def}) as
	\begin{eqnarray}
		    \sum_{i=1}^d p_i{(k)}S(\tilde{\rho}_{\textrm{B}i}^{(k)})
		&=&S\left(\sum_{i=1}^d p_i^{(k)} \Pi_{{\rm A}i}^{(k)}{\otimes}\tilde{\rho}_{\textrm{B}i}^{(k)}\right)
			 -S(\tilde{\rho}_\textrm{A}^{(k)})																											\nonumber \\
		&=&S(\tilde{\rho}^{(k)}_\textrm{AB}) - S(\tilde{\rho}_\textrm{A}^{(k)}).
	\end{eqnarray}
	We recall that $\tilde{\rho}_{\textrm{B}i}^{(k)}$ is the reduced density operator of subsystem B corresponding to the result $i$ after the measurement
	$\mathcal{M}_\textrm{A}^{(k)}$ on subsystem A. $\tilde{\rho}_\textrm{A}^{(k)}$, $\tilde{\rho}_\textrm{B}^{(k)}$ and $\tilde{\rho}_\textrm{AB}^{(k)}$ stand for states of
	subsystems A, B and joint system AB, respectively, after the measurement $\mathcal{M}_\textrm{A}^{(k)}$. Therefore, thermal discord can be expressed as
	\begin{eqnarray}
		D_\textrm{th}^{(k)}(\textrm{B}|\textrm{A})	&=&S(\tilde{\rho}^{(k)}_\textrm{AB})-S(\tilde{\rho}_\textrm{A}^{(k)})
		   +S(\tilde{\rho}_\textrm{A}^{(k)}) - S(\rho_\textrm{AB})						\nonumber\\
		&=&\left\lbrack{S(\tilde{\rho}^{(k)}_\textrm{AB}) - S(\rho_\textrm{B})}\right\rbrack
		   -\left\lbrack{S(\rho_\textrm{AB}) - S(\rho_\textrm{B})}\right\rbrack		\nonumber\\
		&=&S(\mathcal{M}_\textrm{A}^{(k)}|\textrm{B}) - S(\textrm{A}|\textrm{B}),
		   \label{discord_coherence_increaseOfConditionalInfor}
	\end{eqnarray}
	where we have used the fact that the projective measurements are performed
 on subsystem A leading to the equality, ${\rm Tr}_\textrm{A}\tilde{\rho}_\textrm{AB}^{(k)} = {\rm Tr}_\textrm{A}\rho_\textrm{AB} = \rho_\textrm{B}$.
It is now straightforward to
	rewrite Eq.~(\ref{Multi-Meas_UP_Memory}) as
	\begin{equation}
		\sum_{k=1}^N D_\textrm{th}^{(k)}\left(\textrm{B}|\textrm{A}\right)
		{\ge} -\log _2{b} - S\left(\textrm{A}|\textrm{B}\right),
		\label{CR_ThermalDiscord}
	\end{equation}
	which serves as a complementary relation for thermal discord with respect to different measurement bases. This lower bound contains a term of conditional information of the
	pre-measurement state, and thus is also state dependent. As we have mentioned the negativity of conditional information signals entanglement,  for a more entangled state, we
	can expect higher thermal discord. We remark that the inequality (\ref{CR_ThermalDiscord}), resulting from the entropic uncertainty relation with the assistance of
a quantum memory, implies that  thermal discord as a resource for different bases is larger than a bound depending  on both measurement bases and the bipartite state.
It will be interesting if applications of this inequality can be found in statistical mechanics \cite{zurek2003quantum}.

	One would notice that the relation (\ref{CR_ThermalDiscord}) would  reduce to (\ref{CR_RE-Coherence}) when the dimension of the Hilbert space of B is reduced to $1$. This indicates a close
	relation between the relative entropy coherence and thermal discord. In this case, thermal discord would reduce to a relative entropy coherence measure, and conditional information
	$S(\textrm{A}|\textrm{B})$ would equal to $S(\rho_\textrm{A})$ since $\rho_\textrm{AB}=\rho_\textrm{A}$ and $\rho_\textrm{B}=1$.

%====================================================================================================================================================================================
%	Complementary relation of post-measurement conditional information
%====================================================================================================================================================================================
\subsection{Complementary relation of post-measurement conditional information}
Next, we consider a multipartite system, including $N+2$ parties $\textrm{A}\textrm{B}_0{\cdots}\textrm{B}_N$.
We would consider $N+1$ measurements $\mathcal{M}^{(k)}_\textrm{A}$ on the subsystem A. It means the number
of multiple measurements performed on A is equal to the number of subsystems $B_0B_1...B_N$.
	By introducing an ancillary subsystem C, we can always purify
	$\rho_{\textrm{A}\textrm{B}_0\textrm{B}_k}$ to $\rho_{\textrm{A}\textrm{B}_0\textrm{B}_k\textrm{C}}=
(|\Psi \rangle \langle \Psi |)_{\textrm{A}\textrm{B}_0\textrm{B}_k\textrm{C}}$, satisfying $\rho_{\textrm{A}\textrm{B}_0\textrm{B}_k}
={\rm Tr_C}(|\Psi \rangle \langle \Psi |)_{\textrm{A}\textrm{B}_0\textrm{B}_k\textrm{C}}$.
The projective measurement is still performed on subsystem A. Specifically, for the $k$-th measurement
$\mathcal{M}_\textrm{A}^{(k)}$, we can use the corresponding purified state to express the
measurement as,
\begin{eqnarray}
\left( \Pi _{\textrm{A}i}^{(k)}{\otimes}I_{\textrm{B}_0\textrm{B}_k\textrm{C}}\right)|\Psi \rangle _{\textrm{A}\textrm{B}_0\textrm{B}_k\textrm{C}}
=\sqrt{p_i^{(k)}}|i^{(k)}\rangle _{\textrm{A}} |\tilde {\Psi }_i\rangle _{\textrm{B}_0\textrm{B}_k\textrm{C}},
\nonumber \\
\label{schmidt}
\end{eqnarray}
which shows that the post-measurement state of $\textrm{B}_0\textrm{B}_k\textrm{C}$ corresponding to result $i$ of
	measurement $\mathcal{M}_\textrm{A}^{(k)}$ on subsystem A is a pure state with the form $|\tilde {\Psi }_i\rangle _{\textrm{B}_0\textrm{B}_k\textrm{C}}$.
It is known that for the pure state $|\tilde {\Psi }_i\rangle _{\textrm{B}_0\textrm{B}_k\textrm{C}}$,
based on Schmidt decomposition \cite{nielsen2010quantum},
the von Neumann entropy of the two reduced density matrices by
partition $\textrm{B}_0\textrm{B}_k:\textrm{C}$ are the same,
\begin{eqnarray}
 S\left(\tilde{\rho}_{\textrm{B}_0i}\right) = S\left(\tilde{\rho}_{\textrm{B}_k\textrm{C}i}\right).
 \label{2add1}
\end{eqnarray}
In addition, still based on Schmidt decomposition,
 the two density matrices can be transferred to each other by a unitary transformation.
Similarly, let us consider the pure state $|\Psi \rangle _{\textrm{A}\textrm{B}_0\textrm{B}_k\textrm{C}}$, we also know,
\begin{eqnarray}
S\left(\rho_{\textrm{A}\textrm{B}_0}\right) = S\left(\rho_{\textrm{B}_k\textrm{C}}\right).
\label{2add2}
\end{eqnarray}

By taking summation over $i$ for Eq.~(\ref{schmidt}), we have the relation below,
\begin{eqnarray}
&&\sum _i\left( \Pi _{\textrm{A}i}^{(k)}{\otimes}I_{\textrm{B}_0\textrm{B}_k\textrm{C}}\right)|\Psi \rangle _{\textrm{A}\textrm{B}_0\textrm{B}_k\textrm{C}}
\nonumber \\
&&=\sum _i\sqrt{p_i^{(k)}}|i^{(k)}\rangle _{\textrm{A}} |\tilde {\Psi }_i\rangle _{\textrm{B}_0\textrm{B}_k\textrm{C}}.
\label{schmidt1}
\end{eqnarray}
Starting from this state, with the help of the obtained results in Eqs.~(\ref{2add1},\ref{2add2}), we can find,
	\begin{eqnarray}
		&& S(\mathcal{M}_\textrm{A}^{(k)}|\textrm{B}_0) - S(\textrm{A}|\textrm{B}_0)															\nonumber\\
		&=&S\left(\sum_{i=1}^d p_i \Pi_{\textrm{A}i}^{(k)}{\otimes}\tilde{\rho}_{\textrm{B}_0i}^{(k)}\right) - S(\rho_{\textrm{A}\textrm{B}_0})		\nonumber\\
		&=&S\left(\sum_{i=1}^d p_i \Pi_{\textrm{A}i}^{(k)}{\otimes}\tilde{\rho}_{\textrm{B}_k\textrm{C}i}^{(k)}\right) - S(\rho_{\textrm{B}_k\textrm{C}})\nonumber\\
		&\le&S\left(\sum_{i=1}^d p_i \Pi_{\textrm{A}i}^{(k)}{\otimes} \tilde{\rho}_{\textrm{B}_ki}^{(k)}\right) - S(\rho_{\textrm{B}_k})		\nonumber\\
		&=&S(\mathcal{M}_\textrm{A}^{(k)}|\textrm{B}_k),
		\label{Tripartite_InEq_ConditionalInformation}
	\end{eqnarray}
	where the inequality is due to the strong subadditivity of the von Neumann entropy \cite{nielsen2010quantum},
$S\left(\rho_{\textrm{A}\textrm{B}_k\textrm{C}}\right) + S\left(\rho_{\textrm{B}_k}\right)
	{\le}S\left(\rho_{\textrm{A}\textrm{B}_k}\right) + S\left(\rho_{\textrm{B}_k\textrm{C}}\right)$ leading to
	$S\left(\rho_{\textrm{A}\textrm{B}_k\textrm{C}}\right) - S\left(\rho_{\textrm{B}_k\textrm{C}}\right)
	{\le}S\left(\rho_{\textrm{A}\textrm{B}_k}\right) - S\left(\rho_{\textrm{B}_k}\right)$.
Now by taking the summation for $k=1,2,...,N$ on both sides of the inequality (\ref{Tripartite_InEq_ConditionalInformation}), we find,
\begin{eqnarray}
\sum _{k=1}^NS(\mathcal{M}_\textrm{A}^{(k)}|\textrm{B}_0) - NS(\textrm{A}|\textrm{B}_0)
\le \sum _{k=1}^NS(\mathcal{M}_\textrm{A}^{(k)}|\textrm{B}_k).
\nonumber \\
\label{post-inequality}
\end{eqnarray}
On the other hand, with the help of Eq.~(\ref{Multi-Meas_UP_Memory}) where we replace $B$ by
$B_0$, the above inequality (\ref{post-inequality}) leads to,
	\begin{equation}
		S(\textrm{A}|\textrm{B}_0)+\sum_{k=1}^NS(\mathcal{M}_\textrm{A}^{(k)}|\textrm{B}_k)
		{\ge}-\log _2{b}.
	\end{equation}
	Also we know that projective measurements can increase the entropy, and as a consequence, we have,
	\begin{align}
		S(\mathcal{M}_\textrm{A}^{(k)}|\textrm{B}_0) - S(\textrm{A}|\textrm{B}_0)
		=S(\tilde{\rho}_{\textrm{AB}_0}^{(k)}) - S(\rho_{\textrm{AB}_0})				\ge0.
	\end{align}
	Thus, substituting $S(\textrm{A}|\textrm{B}_0)$ by $S(\mathcal{M}_\textrm{A}^{(k)}|\textrm{B}_0)$,
we can further obtain the complementary relation of post-measurements conditional information $S(\mathcal{M}_\textrm{A}^{(k)}|\textrm{B}_k)$ as
	\begin{equation}
		\sum_{k=0}^N S(\mathcal{M}_\textrm{A}^{(k)}|\textrm{B}_k)
		{\ge} -\log _2{b},
		\label{CR_post-measure-ConditionalInformation}
	\end{equation}
  which is similar to Eq.~(\ref{CR_RE-Coherence}) of the relative entropy coherence measure and Eq.~(\ref{CR_ThermalDiscord}) of thermal discord.
One merit of this bound is that it is state
	independent. Also it describes a constraint on bipartite quantum correlations in a multipartite system (usually with more than two
	subsystems), which is quite significant. One can also notice that when $N=1$,
this inequality can reduce to the uncertainty relation in Ref.~\cite{coles2012uncertainty} with two measurements
for a tripartite state.
Our results in this section may stimulate study of quantum correlations concerning multipartite systems.

\section{discussion}
	By utilizing the uncertainty principle formulated in terms of entropies, we explicitly give a group of complementary relations for
	quantum coherence, thermal discord and
	conditional information. Specifically, the entropic uncertainty relation without a quantum memory would give us a lower bound (\ref{CR_RE-Coherence}) on quantum coherence in the system of
	a single particle with multiple measurements. It shows that the summation of the coherence measure of a quantum state with respect to different measurement bases should have a lower
	bound. Also the purer the state which corresponds to a lower entropy, the higher this bound would be.
These entropic uncertainty relations concerning additional memories
	provide a lower bound in Eq.~(\ref{CR_ThermalDiscord})
on thermal discord with respect to different projective measurements on a subsystem of a bipartite joint system. This lower bound
	is for the summation of thermal discord with respect to multiple measurements on the chosen subsystem.
 It indicates that the more entanglement of the state which corresponds to higher
 minus conditional information, $-S({\rm A}|{\rm B})$, the higher this lower bound would be. The latter group of complementary relations about thermal discord can be reduced to that of quantum
	coherence by discarding the auxiliary memory through setting its dimension to $1$.
In addition, we also derive a state-independent lower bound for
	the sum of post-measurement conditional
	information between a specific subsystem and the other subsystems.
These three complementary relations indicate that there are important constraints for quantum quantities with multiple measurements.
The results also imply that there exits a delicate relation between uncertainty principle with quantum coherence and quantum correlations in quantum mechanics.

%====================================================================================================================================================================================
%	Acknowledge
%====================================================================================================================================================================================
\begin{acknowledgments}
	This work was supported by the National Key R \& D Plan of China (No. 2016YFA0302104,
No. 2016YFA0300600), the National Natural Science Foundation of China (Nos. 91536108, 11774406),
and Strategic Priority Research Program of Chinese Academy of
Sciences (Grant No. XDB28000000).
\end{acknowledgments}

\noindent {\bf Author contributions:}
H.F. and L.Z.M. proposed the project. Y.P. and Z.Y.F. made the calculations with assistance from H.F.,
Y.P. wrote the paper with assistance from Z.Y.F. and H.F., the manuscript preparation
was commented by L.S., Y.R.Z. and L.Z.M.

\noindent {\bf Competing interests:} The authors declare that they have no competing interests.

%\bibliography{Bibliography}


\end{document}